\PassOptionsToPackage{pdftex}{graphicx}   % graphicx を pdftex ドライバで読む
\PassOptionsToPackage{update,prepend}{epstopdf} % ← EPS 自動変換したいなら

\documentclass[fp]{jpsj3} % JPSJ format
\usepackage{graphicx} % Include figure files
\usepackage{xcolor}
\newcommand{\rev}[1]{#1}                    % arXiv v2 build: no color (this project)
\usepackage{dcolumn} % Align table columns on decimal point
\usepackage{bm} % bold math
\usepackage{algpseudocode,algorithm}
\usepackage[pagewise]{lineno}
\usepackage{amsmath, amssymb}
\usepackage{here}
\usepackage{subfigure}
\usepackage{txfonts}
\usepackage{epstopdf}   % 自動変換

\usepackage{comment}

\renewcommand{\figurename}{Fig.}

\title{Automatic Termination Strategy of Inelastic Neutron-scattering Measurement Using Bayesian Optimization for Bin-width Selection}

\author{Kensuke~Muto$^{1,4}$,
        Hirotaka~Sakamoto$^1$,
        Kenji~Nagata$^2$,
        Taka\mbox{-}hisa~Arima$^3$,
        and Masato~Okada$^{2,3}$\thanks{okada@edu.k.u-tokyo.ac.jp}}

\inst{$^1$Graduate School of Frontier Sciences, The University of Tokyo, Chiba 277--0882, Japan \\
      $^2$Research and Services Division of Materials Data and Integrated System, National Institute for Materials Science, Tsukuba, Ibaraki 305--0047, Japan \\
      $^3$Graduate School of Frontier Sciences, The University of Tokyo, Chiba 277--8561, Japan \\
      $^4$Karakuri, Inc., 2-7-3 Tsukiji, Chuo-ku, Tokyo 104-0045, Japan}

\abst{
Currently, an excessive amount of event data is being obtained in four-dimensional inelastic neutron-scattering experiments.
A method for automatic bin-width optimization of multidimensional histograms has been developed and recently validated on real inelastic neutron-scattering data.
However, measuring beyond the equipment resolution leads to inefficient use of valuable beam time.
To improve experimental efficiency, an automatic termination strategy is essential.
\rev{We propose a Bayesian-optimization-based method to compute a stopping criterion that can support online decisions on whether to continue or terminate an experiment.}
In the proposed method, the bin-width optimization is performed using Bayesian optimization to efficiently compute the optimal bin widths.
The experiment is terminated when the optimal bin widths become smaller than the target resolutions.
In numerical experiments using real inelastic neutron-scattering data, the optimal bin widths decrease as the number of events increases.
Even the optimal bin widths for data downsampled to $1/5$ are comparable with the resolutions limited by the sample size, choppers, and so on.
This implies excessive measurement of the inelastic neutron experiments for the moment.
Moreover, we found that Bayesian optimization can reduce the search cost to approximately 10\% of an exhaustive search in our numerical experiments.
}

\begin{document}
\maketitle
% \linenumbers % [v7#1] disabled in this arXiv v2 project (JPSJ referee build only)

\section{Introduction\label{sec:introduction}}
Inelastic neutron-scattering is an experimental method for investigating the dynamical structure of materials \cite{neutron_scattering_tech}.
In recent years, numerous event data have been obtained in inelastic neutron-scattering experiments using high-power accelerator-based neutron sources such as those of ISIS, \cite{thomason2019isis}
SNS, \cite{mason2006spallation}
J-PARC, \cite{takada2017materials}
and CSNS \cite{chen2025design}.
A time-of-flight neutron spectrometer designed for inelastic scattering measurements, such as
MAPS \cite{MAPS},
MERLIN \cite{MERLIN},
HYSPEC \cite{HYSPEC},
ARCS \cite{ARCS},
4SEASONS \cite{4SEASONS},
AMATERAS \cite{AMATERAS}, and
HRC \cite{HRC},
produces a large-scale event data.
Researchers obtain event data mapped on the four-dimensional (4D) space of transferred energy ($E$) and momentum ($\bf{q}$).
At present, researchers create histograms from the obtained event data and analyze them \cite{Utsusemi}.
It is necessary to set the bin widths when creating a histogram.

Shimazaki and Shinomoto proposed a one-dimensional histogram bin-width optimization method based on minimizing a cost function, originally developed for neuronal spike trains in neuroscience \cite{Shimazaki}.
They also proposed an extrapolation theory to predict how optimal bin width changes as the number of trials increases.
This extrapolation theory provides a practical way to estimate how many additional experimental trials are needed to construct a meaningful histogram with sufficient resolution.
Muto et al.\ extended this approach to multidimensional histograms by using summed-area tables (SAT) \cite{Muto,SAT}.
The SAT implementation substantially reduces the computational cost of evaluating the cost function for multidimensional histograms.
Recently, Tatsumi et al.\ applied this bin-width optimization method to real inelastic neutron-scattering data and validated both the optimization method and the extrapolation theory for Cu single crystal measurements at J-PARC \cite{tatsumi2022optimization}.

\begin{figure}[ht]
    \begin{center}
		 \includegraphics[width = \linewidth]{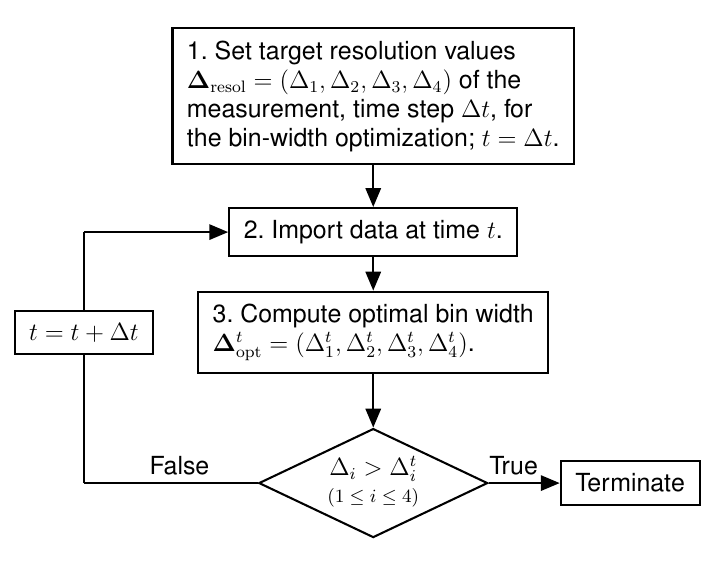}
    \end{center}
%	\vspace{-0.3cm}
    \caption{Overview of the termination strategy.}
    \label{fig:concept}
\end{figure}

A practical application of multidimensional bin-width optimization is to determine whether to stop or to continue the experiment.
By using a larger amount of data, we can extract more detailed features of the target materials.
However, measuring beyond the resolution of the measurement equipment leads to inefficient use of valuable beam time.
% [v7#3 referee(2)] original:
% While the extrapolation theory predicts how optimal bin widths change with data size, it cannot exactly predict the optimal bin widths for each specific measurement; thus, accurate stopping decisions require real-time bin-width optimization during the experiment.
\rev{While the extrapolation theory predicts how optimal bin widths change with data size, it cannot exactly predict the optimal bin widths for each specific measurement; thus, accurate stopping decisions require bin-width optimization during the experiment.}
% [v7#4 referee(2)] original (4 lines -> 3 lines; the operational-resources sentence is removed):
% Tatsumi et al.\ demonstrated that parallel computation on a 32-core Xeon processor enabled such real-time optimization \cite{tatsumi2022optimization}.
% However, parallel computation infrastructure requires significant operational resources.
% We propose a Bayesian-optimization-based method to compute the stopping criteria and determine whether to stop or continue the experiment in real time. The flowchart of the method is shown in Fig. \ref{fig:concept}.
% By using Bayesian optimization to efficiently search for optimal bin widths, our approach achieves real-time performance without requiring parallel-computation infrastructure.
\rev{Tatsumi et al.\ demonstrated rapid bin-width optimization using a parallelized Fortran implementation with MPI on a 32-core Xeon processor \cite{tatsumi2022optimization}. Here, we investigate Bayesian optimization as an alternative strategy for reducing the number of evaluated bin-width candidates.}
\rev{We propose a Bayesian-optimization-based method to compute a stopping criterion that can support online decisions on whether to stop or continue an experiment.} The flowchart of the method is shown in Fig. \ref{fig:concept}.
\rev{By efficiently searching for the optimal bin widths, the proposed method reduces the number of evaluated bin-width candidates without requiring an explicit MPI-based parallel implementation.}
The experiment is terminated when the optimal bin widths become smaller than the expected resolutions.
We conducted numerical experiments using real experimental data and found that the optimal bin widths decrease as the number of data increases, which is consistent with previous studies \cite{Shimazaki, Muto, tatsumi2022optimization}.
Moreover, we found that Bayesian optimization can reduce the search cost to approximately 10\% of exhaustive search in our numerical experiments.
% [v7#5 referee(2)] original:
% This computational efficiency demonstrates that our approach can serve as a practical stopping criterion for routine inelastic neutron-scattering experiments.
\rev{This reduction in search cost indicates the potential of our approach to support online stopping decisions in inelastic neutron-scattering experiments.}

The structure of this paper is as follows. In Section \ref{sec:method}, we formulate a method of multidimensional bin-width optimization and Bayesian optimization.
In Section \ref{sec:results}, we present the numerical experiments using some real experimental data to explain the significance of studying a stopping strategy and efficiency of Bayesian optimization.
We discuss the results in Section \ref{sec:discussions} and provide conclusions in Section \ref{sec:conclusion}.

\section{Method\label{sec:method}}
\subsection{Bin-width optimization for one-dimensional event data}
Shimazaki and Shinomoto proposed a bin-width optimization method for 1D data \cite{Shimazaki}. They considered a case where researchers obtain $n$ pieces of event data $t_i \in [0, T] \ (1 \le i \le n)$ in the observation time interval $[0, T]$ and the data follow a true probability density $\lambda(t)$. The performance of the estimator $\hat{\lambda}(t)$ for $\lambda(t)$ can be evaluated using the mean integrated squared error (MISE) as follows,
\begin{eqnarray}
{\rm MISE} = \frac{1}{T} \int_{0}^{T} E\left[ \left(\hat{\lambda} (t) - \lambda(t) \right)^2 \right]dt.
\label{eq:MISE}
\end{eqnarray}
In statistics, MISE is used for density estimations \cite{MISE}. Here, $E[\cdot]$ represents the expectation over different realizations of event data given by $\lambda(t)$. When the observation interval is equally divided into $N$ parts, the estimator $\hat{\lambda}(t)$ is limited to the form of the histogram of the bin width $\Delta = \frac{T}{N}$. A cost function is derived by extracting the terms depending on the bin width $\Delta$ from the right side of Eq. (\ref{eq:MISE}), as
\begin{eqnarray}
\hat{C}(\Delta) &=& \frac{2\bar{k} - v}{\Delta^2}, \\
{\rm where} \	 \bar{k} &=& \frac{n}{N} , \ v = \frac{1}{N}\sum_{i = 1}^{N} (k_i - \bar{k})^2.
\label{eq:k_i_and_v_1D}
\end{eqnarray}
Here, $k_i$ represents the number of events contained in the $i-$th bin. We define the optimal bin width as the bin width that minimizes the cost function.

\subsection{Multidimensionalization of the bin-width optimization algorithm}
We will consider a case where researchers obtain $n$ pieces of event data mapped on $\mathbb{R}^d$. To make a histogram, we divide the observation area equally into $N$ $d$-dimensional rectangular parallelepipeds with bin widths $\Delta_{1}, \Delta_{2}, \dots, \Delta_{d}$. Similarly to the 1D case, we can compute the cost function $\hat {C} _n (\Delta_{1}, \Delta_{2}, \dots , \Delta_{d})$ as
\begin{eqnarray}
\hat{C}(\Delta_{1}, \Delta_{2}, ... , \Delta_{d}) &=& \frac{2\bar{k} - v}{(\Delta_{1} \Delta_{2} ... \Delta_{d})^2}, \\
{\rm where} \	\	\	\	 \bar{k} &=& \frac{n}{N} ,\ v = \frac{1}{N}\sum_{i = 1}^{N} (k_i - \bar{k})^2,
\label{eq:cost_func}
\end{eqnarray}
where $k_i$ represents the number of events contained in the $i-$th bin. The estimator of the height of the $i-$th bin is $\hat{\theta}_i = \frac{k_i}{n \Delta_1 \Delta_2 ... \Delta_d} $. Moreover, in the multidimensional case, optimal bin widths minimize the cost function.

\begin{figure*}[ht]
\begin{center}
 \includegraphics[width = \linewidth]{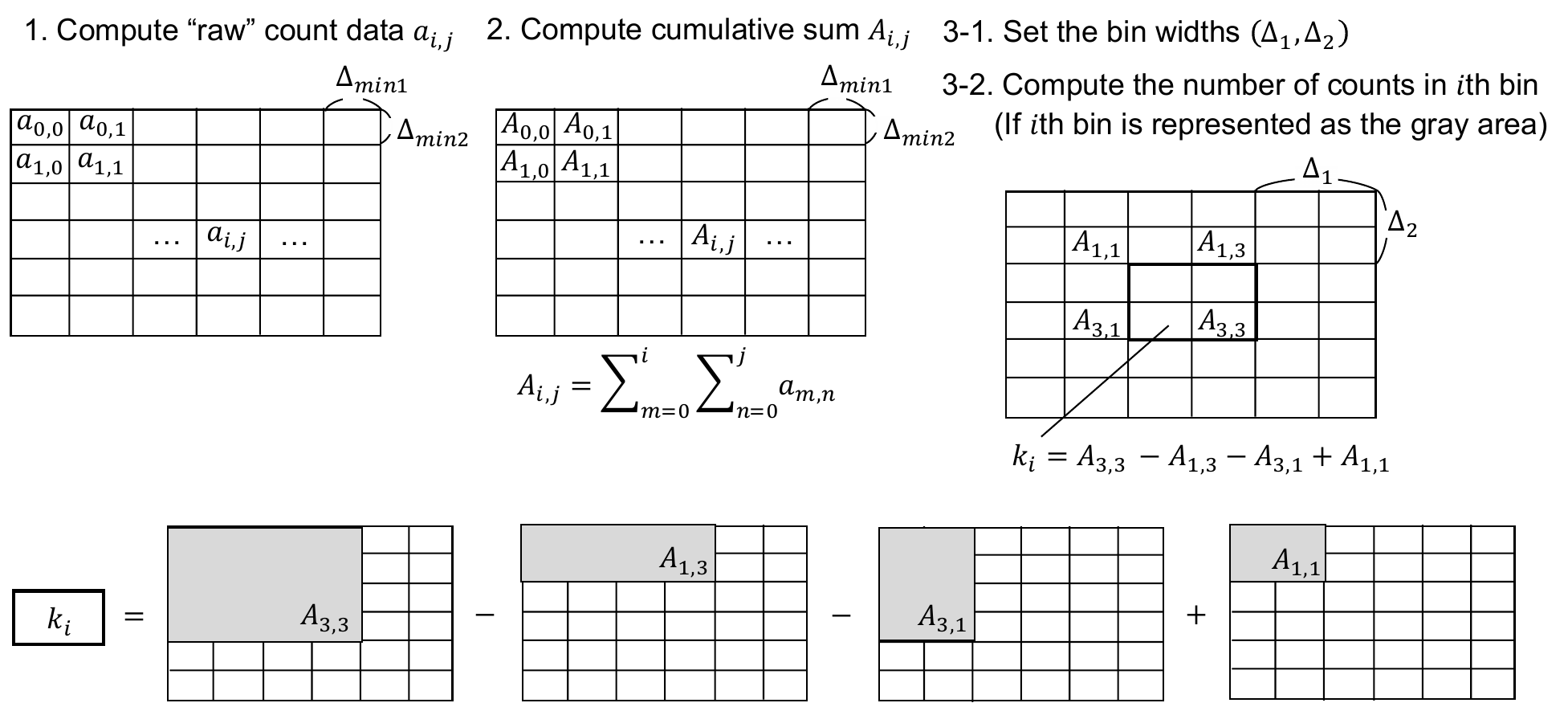}
\end{center}
\caption{ Overview of the summed-area tables algorithm for the 2D case. We set a sufficient fine histogram and call it ``raw'' count data $a_{i,j}$. }
\label{fig:SAT}
\end{figure*}

The computational cost of calculating the optimal bin widths increases as the dimension $d$ increases, and therefore it is essential to reduce the cost. Using the summed-area table (SAT) algorithm, as shown in Fig. \ref{fig:SAT} for $d = 2$, we can reduce the computational complexity for the number of counts in each bin \cite{SAT}. To apply the SAT algorithm, it is necessary to obtain event data in a sufficiently fine histogram.
The steps of SAT are shown in Algorithm \ref{alg:alg1}. Since it is not necessary for the termination strategy to find the exact optimal bin widths, we consider only multiples of the minimum units of bin widths. It is sufficient to set the maximum value of $\Delta_{i}$ to being several times as large as the required resolution.
\begin{algorithm}[h]
\caption{Bin-width optimization (multidimensional case)}
 \label{alg:alg1}
\noindent Step 1:
 \begin{itemize}
 \setlength{\leftskip}{0.5cm}
\item[{}] Set the minimum units of the bin widths: $\Delta_{min, 1}, \Delta_{min, 2}, ... , \Delta_{min, d}$ and compute ``raw'' count data as shown in Fig. \ref{fig:SAT}(a). Compute the cumulative sum $A_{i_1, i_2, ..., i_d}$ of the ``raw'' count data for preparation of the SAT algorithm.
 \end{itemize}
\noindent Step 2:
 \begin{itemize}
 \setlength{\leftskip}{0.5cm}
\item[{}] Set the bin widths $(\Delta_{1}, \Delta_{2}, ... , \Delta_{d})$. Compute the number of counts $k_i$ in the $i-$th bin by using the SAT algorithm. Then, compute $v$ defined in Eq. (\ref{eq:cost_func}). Compute $\hat{C}(\boldsymbol{\Delta})$ defined in Eq. (4).
 \end{itemize}
\noindent Step 3:
 \begin{itemize}
 \setlength{\leftskip}{0.5cm}
\item[{}] Repeat step 2 to find the bin width $\boldsymbol{\Delta}$ that minimizes $\hat {C}(\boldsymbol{\Delta})$, and let this be the optimal bin width.
 \end{itemize}
\end{algorithm}

Let us discuss the computational complexity of the bin-width optimization. Denote the maximum division number in the $i-$th dimension as $N_{max, i}$. Let $N_i$ be the number of axial divisions in the $i-$th dimension, and consider the case where the cost function is computed by setting $N_i$ from 2 to $N_{max, i}$. Then, the computational cost of the bin-width optimization is $O(\prod_{i=1}^d N_{max, i} \log N_{max, i})$.

\subsection{Bayesian optimization}
Bayesian optimization is a method for solving optimization problems involving an objective function. Moreover, a Gaussian process (GP) is often used to interpolate the objective function by using correlation based on the distance between the observation points. Bayesian optimization uses the results of the interpolation to select the next point to maximize the acquisition function. Here, we treat the expected improvement (EI) \cite{EI} as the acquisition function:
\begin{eqnarray}
EI({\bf x}) = \begin{cases}
(f^* - \mu({\bf x})) \Phi \left(\frac{f^* - \mu({\bf x})}{\sigma({\bf x})}\right) + \sigma({\bf x}) \phi \left(\frac{f^* - \mu({\bf x})}{\sigma({\bf x})}\right), & (\sigma({\bf x}) > 0) \\
0. & (\sigma({\bf x}) = 0)
\end{cases}
\label{eq:EI}
\end{eqnarray}
$\mu({\bf x})$ and $\sigma({\bf x})$ are the mean and variance functions of the Gaussian process at point ${\bf x}$, respectively, and $f^*$ is the minimum value among the observed data. $\phi$ and $\Phi$ represent the normalized Gaussian function and its cumulative distribution function. EI has been proven to have convergence properties \cite{EI_convergence}. The computational complexity of the acquisition function is $O(N^2)$ when the hyperparameter (HP) of GP is fixed, and $O(N^3)$ when the HP of GP is tuned sequentially. Here, $N$ represents the number of searched points.

\section{Results\label{sec:results}}
We performed numerical experiments with experimental data to verify the validity of the proposed method. The data consisted of inelastic neutron-scattering measurements on Ba$_3$Fe$_2$O$_5$Cl$_2$. The unit cell of this material is body-centered cubic with a lattice constant of $9.96 \mathrm{\AA}$\cite{leib1985neuer}. First, we computed the cost function for the whole experimental data and found the optimal bin widths. Then, we conducted downsampling to vary the number of events and computed the cost function for each downsampled data. Then, we examined the relationship between the number of events and the optimal bin widths. Finally, we examined the effectiveness of BO in finding the optimal bin widths. The resolution of the experimental equipment was $\Delta_{\mathrm{resol}} = (5, 0.25, 0.25, 0.30) {\mathrm{[meV, r.l.u., r.l.u., r.l.u.]}}$.

\subsection{Bin-width optimization for experimental data}
\begin{figure*}[t]
	\begin{minipage}{0.325\hsize}
		\begin{center}
			 \subfigure[]{
		 \includegraphics[width = 6cm]{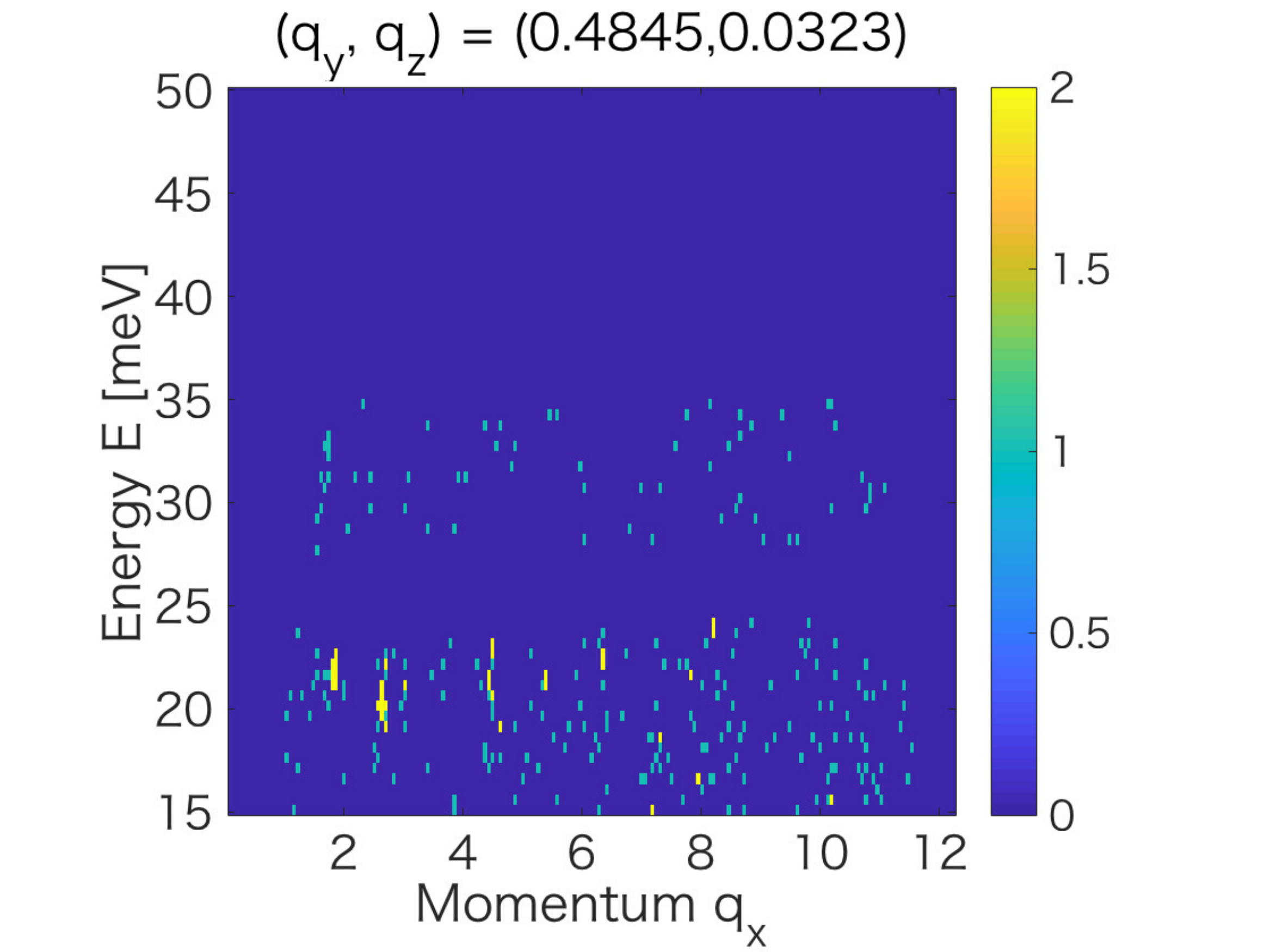}}
		\end{center}
	\end{minipage}
	\begin{minipage}{0.325\hsize}
		\begin{center}
			 \subfigure[]{
				 \includegraphics[width = 6cm]{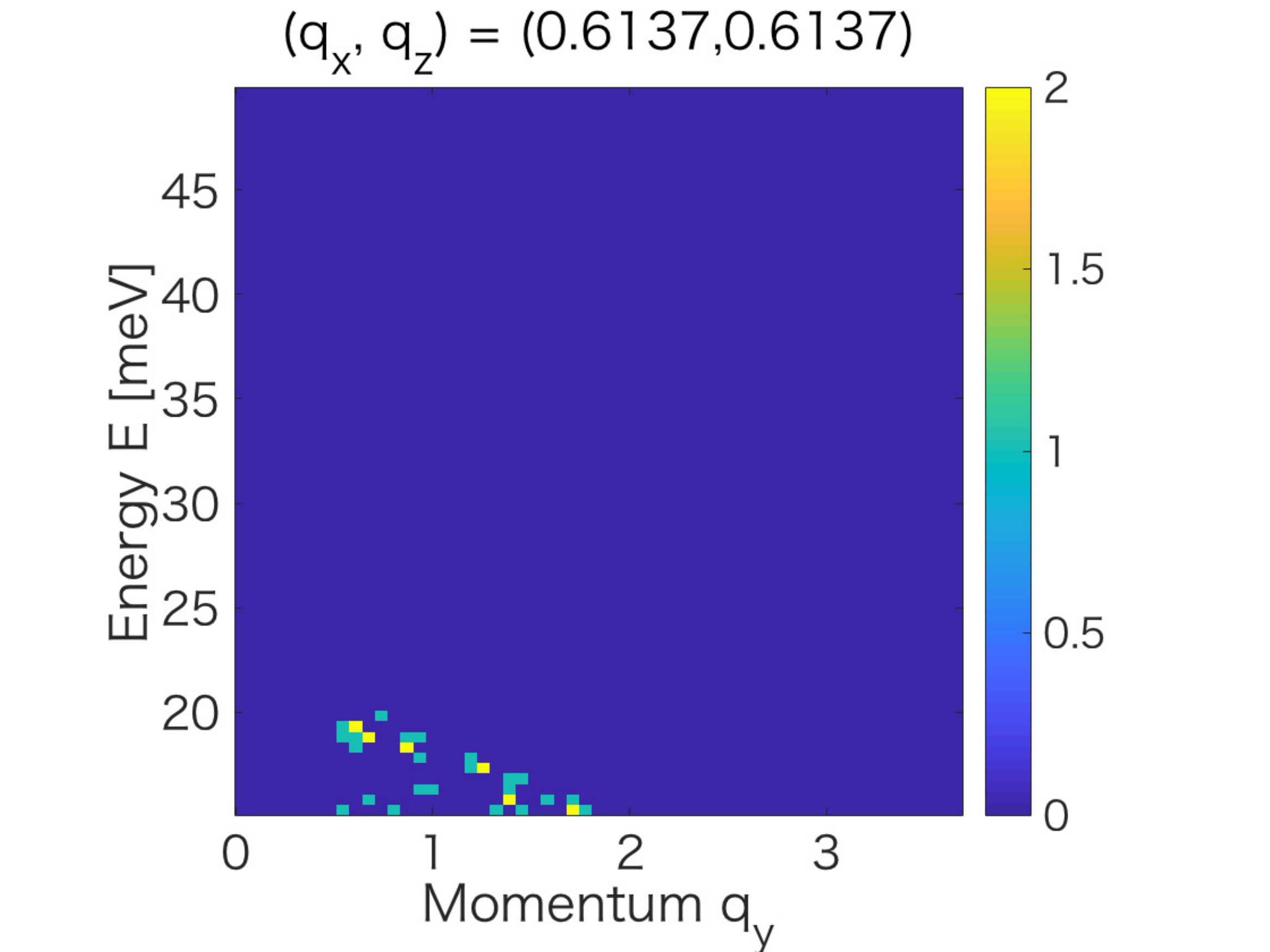}}
		\end{center}
	\end{minipage}
	\begin{minipage}{0.325\hsize}
		\begin{center}
			 \subfigure[]{
				 \includegraphics[width = 6cm]{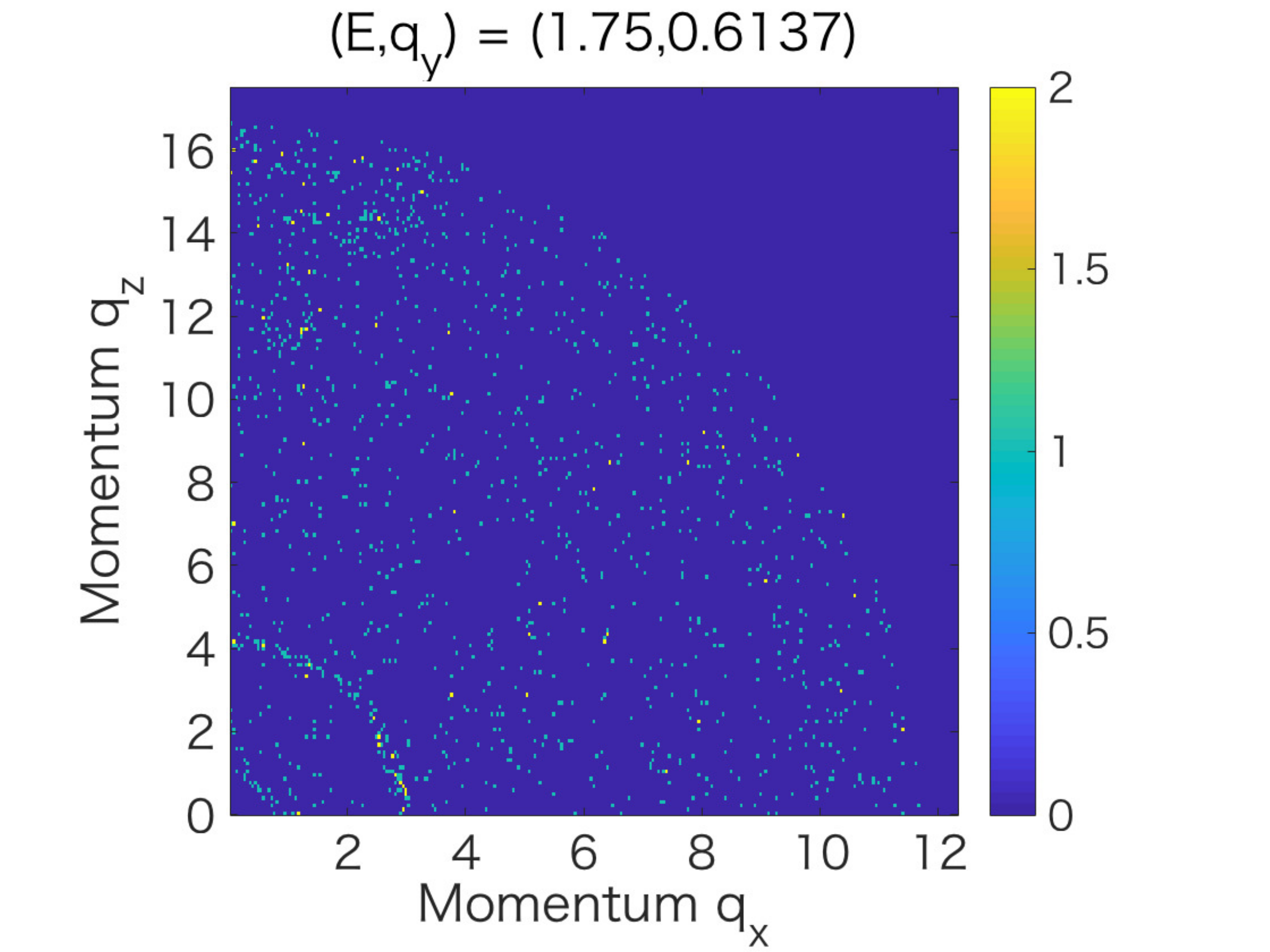}}
		\end{center}
	\end{minipage}
	\vspace{-0.4cm}
\caption{2D slices of 4D inelastic neutron-scattering count data. Regarding (a), $q_y$ and $q_z$ are fixed at $0.4845 \pm 0.0323$ and $0.0323 \pm 0.0323$, respectively. Regarding (b), $q_x$ and $q_z$ are fixed at $0.6137 \pm 0.0323$ and $0.6137 \pm 0.0323$, respectively. Regarding (c), $E$ and $q_y$ are fixed at $1.75 \pm 0.25$ and $0.6137 \pm 0.0323$, respectively. The upper bound of the color axes is set to 2.}
 \label{fig:raw_count_data}
\end{figure*}

\begin{figure*}[h!]
\begin{minipage}{0.45\hsize}
	\begin{center}
		 \subfigure[]{
			 \includegraphics[width = 7cm]{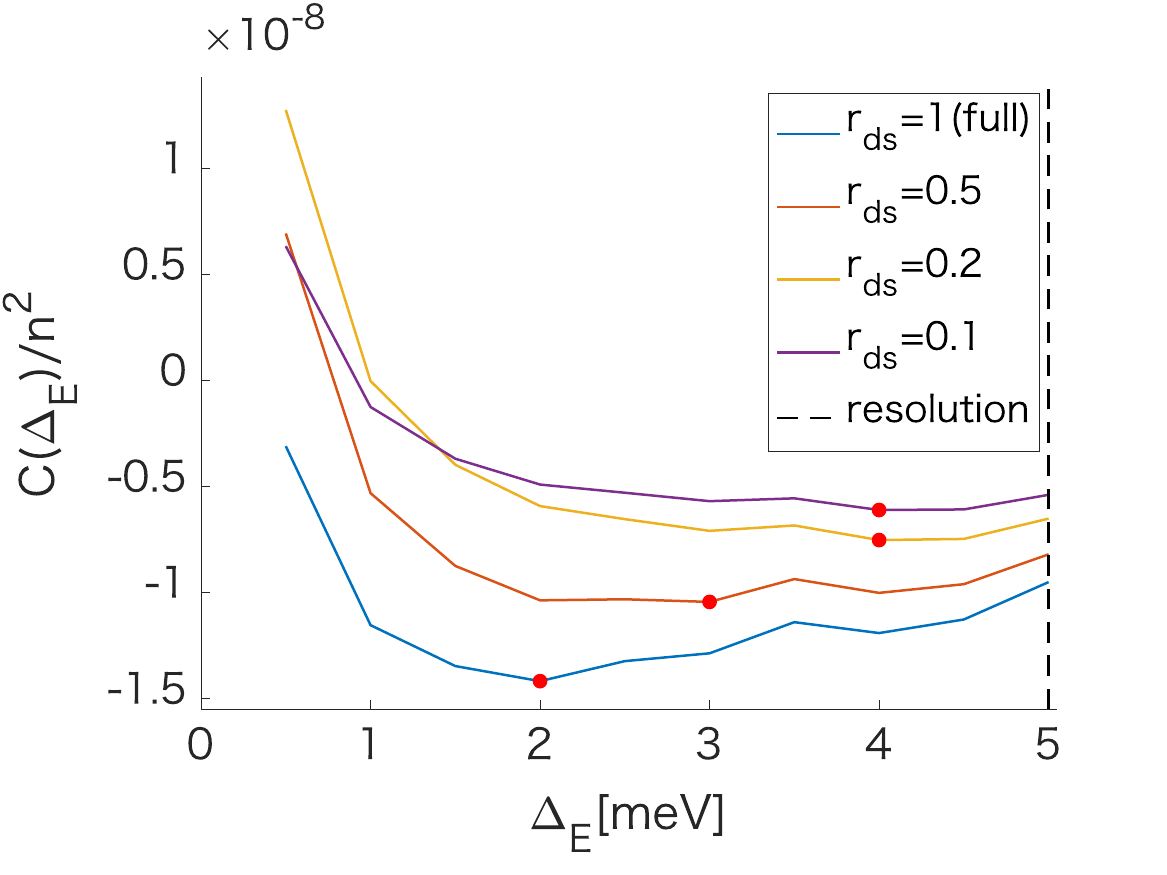}}
	\end{center}
\end{minipage}
\begin{minipage}{0.45\hsize}
	\begin{center}
		 \subfigure[]{
			 \includegraphics[width = 7cm]{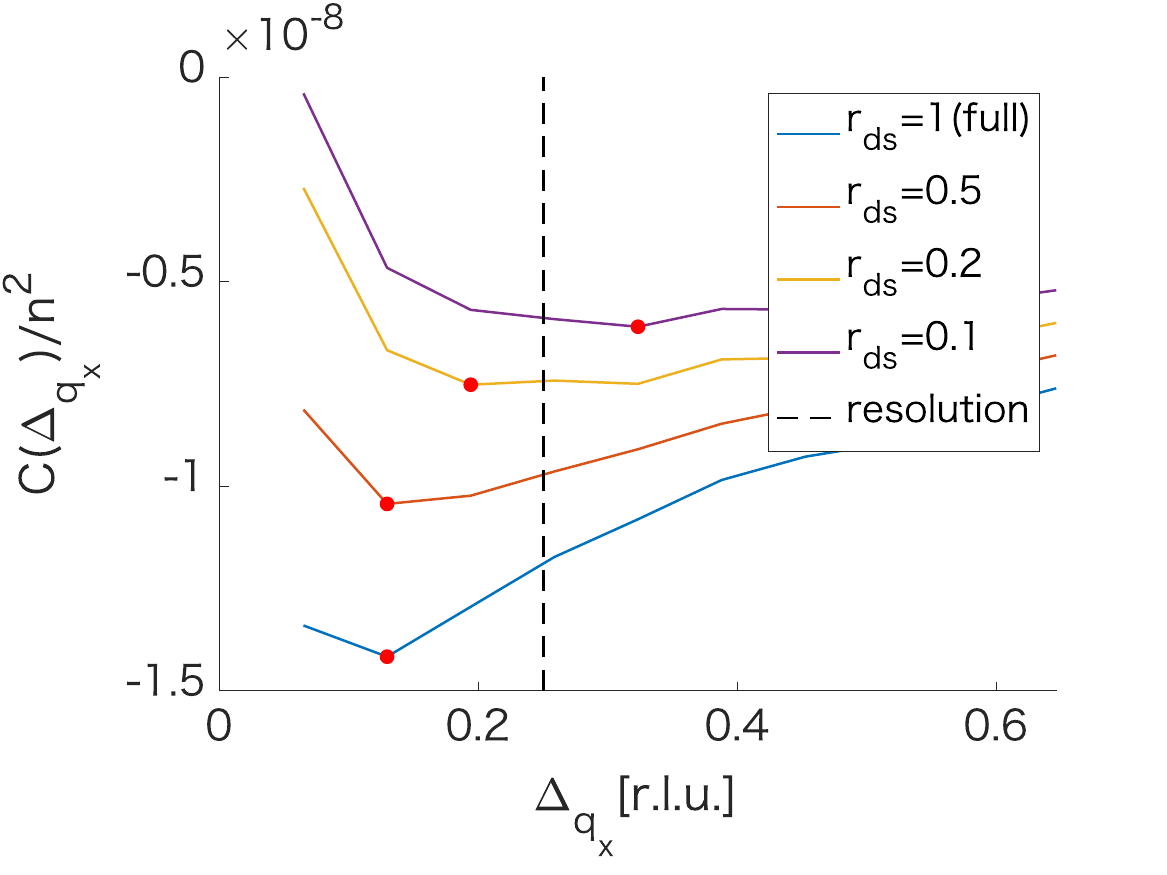}}
	\end{center}
\end{minipage}
\begin{minipage}{0.45\hsize}
	\begin{center}
		 \subfigure[]{
			 \includegraphics[width = 7cm]{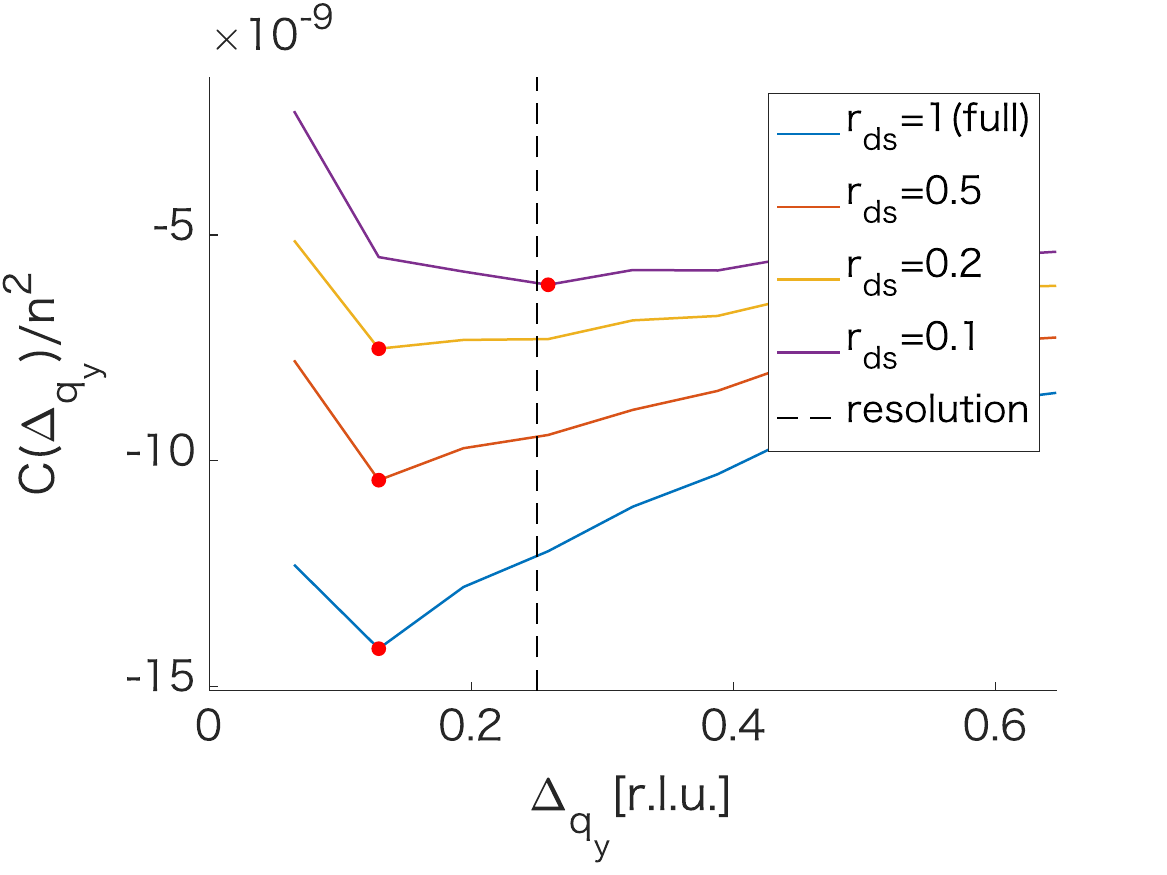}}
	\end{center}
\end{minipage}
\begin{minipage}{0.45\hsize}
	\begin{center}
		 \subfigure[]{
			 \includegraphics[width = 7cm]{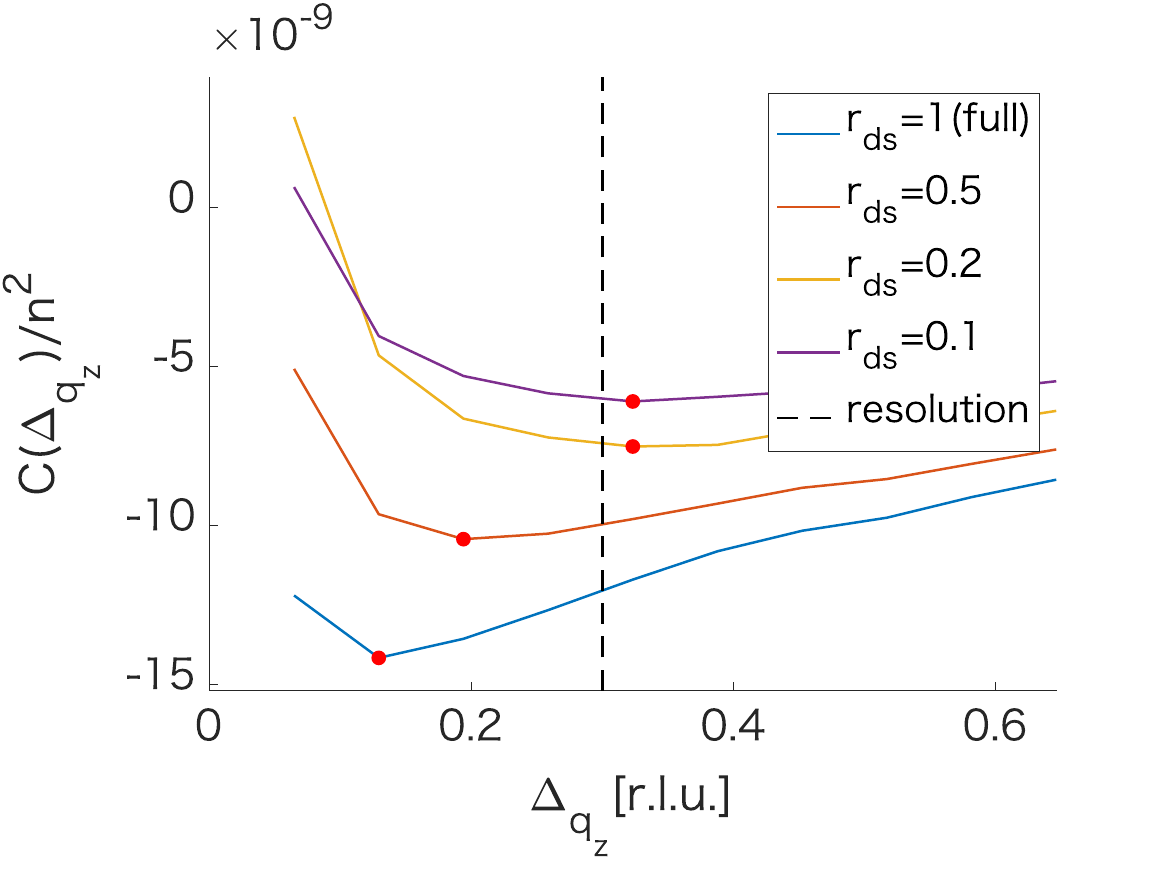}}
	\end{center}
\end{minipage}
\caption{ 1D slices of 4D cost functions computed by changing the downsampling rate of $r_{ds}$. To unify the scale, each cost function is divided by the square of the number of events. (a)--(d) represent landscapes of the cost functions in the $E$, $q_x$, $q_y$, and $q_z$ direction. The cost value in one axis direction is visualized, and the other three axis directions are fixed to the optimal bin width. The optimal bin widths $(\Delta_E, \Delta_{q_x}, \Delta_{q_y}, \Delta_{q_z})$ are $(4{\rm[meV]}, 0.32, 0.26, 0.32)$ for $r_{ds} = 0.1$, $(4{\rm[meV]},0.19, 0.13, 0.32)$ for $r_{ds} = 0.2$, $(3{\rm[meV]},0.13, 0.13, 0.19)$ for $r_{ds} = 0.5$, and $(2{\rm[meV]},0.13, 0.13, 0.13)$ for $r_{ds} = 1$. Approximate optimal bin widths and the experiment's resolutions are represented as filled circles and dashed lines. }
\label{fig:cost_func}
\end{figure*}

We experimented with inelastic neutron-scattering data within $ 15 ~{\rm [meV]} \le E\le 50~ {\rm [meV]}$ and momentum components $0 \le q_x, q_y, q_z$, in which there were $1,679,542$ events.
% [v7#6+#7 self-correction] original ("0.13" was a misstated minimum unit; the three-origin averaging below was already part of the implementation):
% To use the SAT algorithm, we set the minimum units of the bin widths in the energy and momentum directions as $\Delta_{min, E} = 0.5 ~{\rm [meV]}$ and $\Delta_{min, q} = 0.13 ~ {\rm [r.l.u.]}$, respectively. We limited the search area for the optimal bin widths to multiples of the minimum unit. Since we do not consider the case where the number of events is extremely small, we limit the search space of the bin width up to $10$ times the minimum unit in each dimension. Since we considered a 4D space, there were a total of $10,000$ search candidates. To simulate the process of increasing data while conducting a measurement, we generated data downsampled at a rate of $r_{ds}$ from the full data.
To use the SAT algorithm, we set the minimum units of the bin widths in the energy and momentum directions as $\Delta_{min, E} = 0.5 ~{\rm [meV]}$ and \rev{$\Delta_{min, q} = 0.0646 ~ {\rm [r.l.u.]}$}, respectively. We limited the search area for the optimal bin widths to multiples of the minimum unit. Since we do not consider the case where the number of events is extremely small, we limit the search space of the bin width up to $10$ times the minimum unit in each dimension. Since we considered a 4D space, there were a total of $10,000$ search candidates. \rev{For each candidate set of bin widths, the cost function was evaluated at three origins of the histogram grid, shifted simultaneously along all four axes by $0$, $\Delta_i/4$, and $\Delta_i/2$, and the three values were averaged to reduce the sensitivity of the estimate to the choice of the bin origin.} To simulate the process of increasing data while conducting a measurement, we generated data downsampled at a rate of $r_{ds}$ from the full data. We computed the cost functions for $r_{ds} = 0.1, 0.2, 0.5, 1$. Fig. \ref{fig:raw_count_data} shows 2D slices of the 4D data. To visualize the relation between the number of events and the optimal bin widths, we plotted the one-dimensional profile of the cost functions (Fig. \ref{fig:cost_func}). To unify the scale of the cost functions for different data sizes, we divided each cost function by the square of the number of events. The optimal bin widths $(\Delta_E, \Delta_{q_x}, \Delta_{q_y}, \Delta_{q_z})~ {\rm [meV, r.l.u., r.l.u., r.l.u.]}$ were $(4, 0.32, 0.26, 0.32)$, $(4, 0.19, 0.13, 0.32)$, $(3, 0.13, 0.13, 0.19)$, and $(2 ,0.13, 0.13, 0.13)$ for $r_{ds} =$ 0.1, 0.2, 0.5, and 1, respectively. The optimal energy bin width for $r_{ds} = 0.1$ was smaller than the energy resolution of the experiment $\Delta_{\rm resol, E}$.

\begin{figure*}[h!]
\begin{minipage}{0.45\hsize}
	\begin{center}
		 \subfigure[]{
			 \includegraphics[width = 7cm]{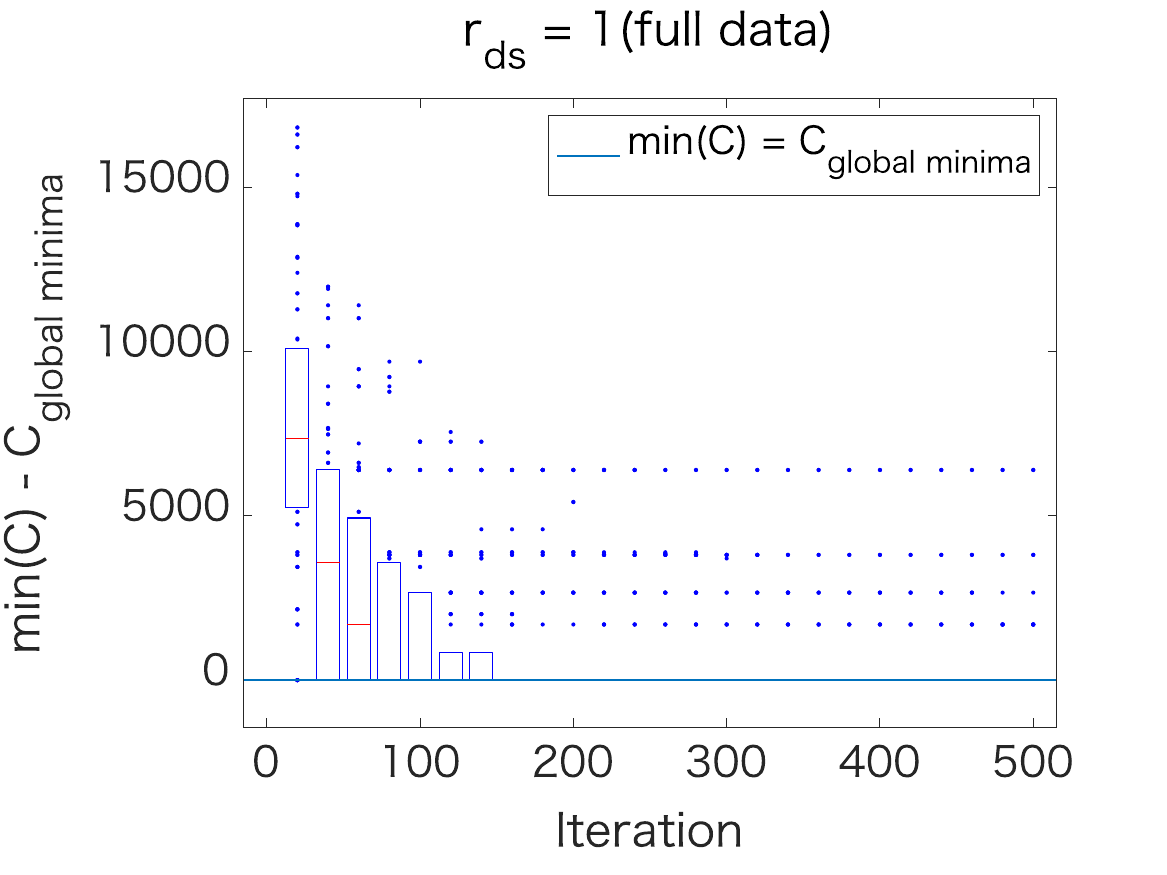}}
	\end{center}
\end{minipage}
\begin{minipage}{0.45\hsize}
	\begin{center}
		 \subfigure[]{
			 \includegraphics[width = 7cm]{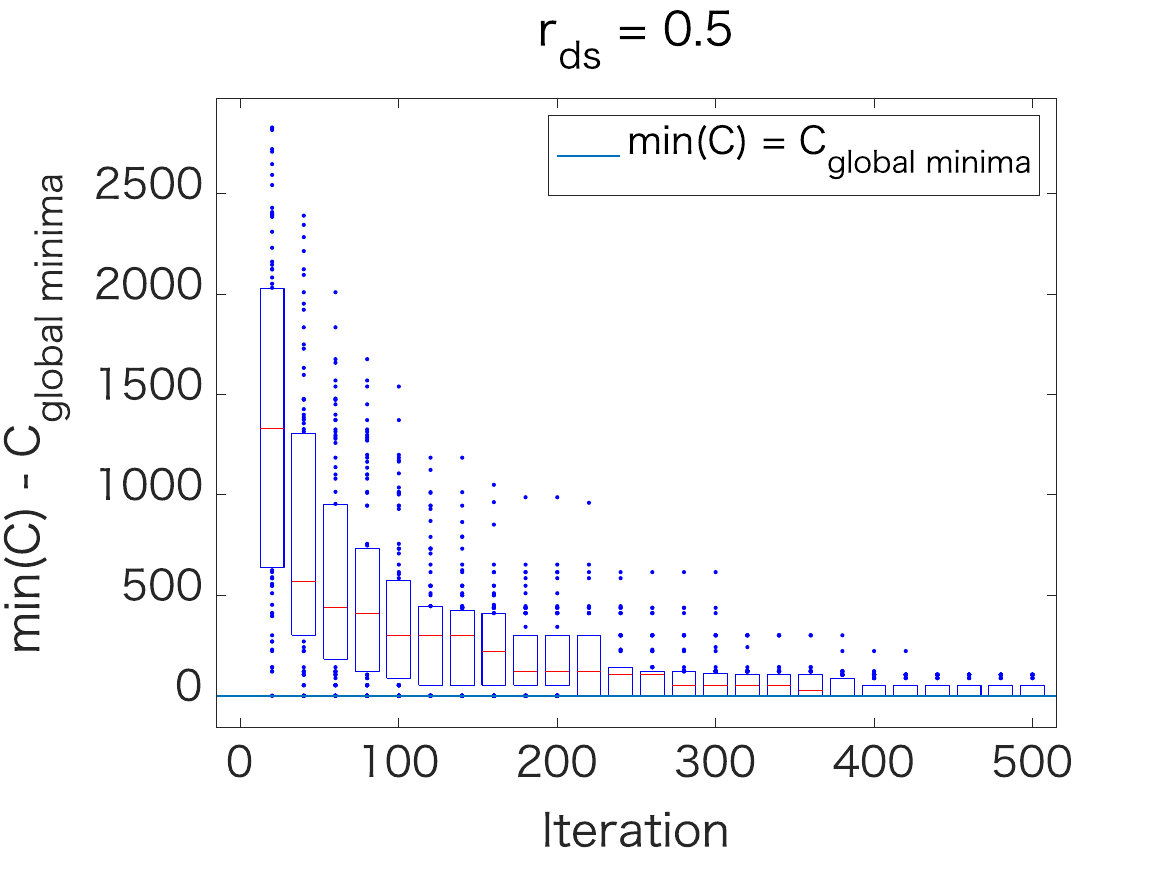}}
	\end{center}
\end{minipage}
\begin{minipage}{0.45\hsize}
	\begin{center}
		 \subfigure[]{
			 \includegraphics[width = 7cm]{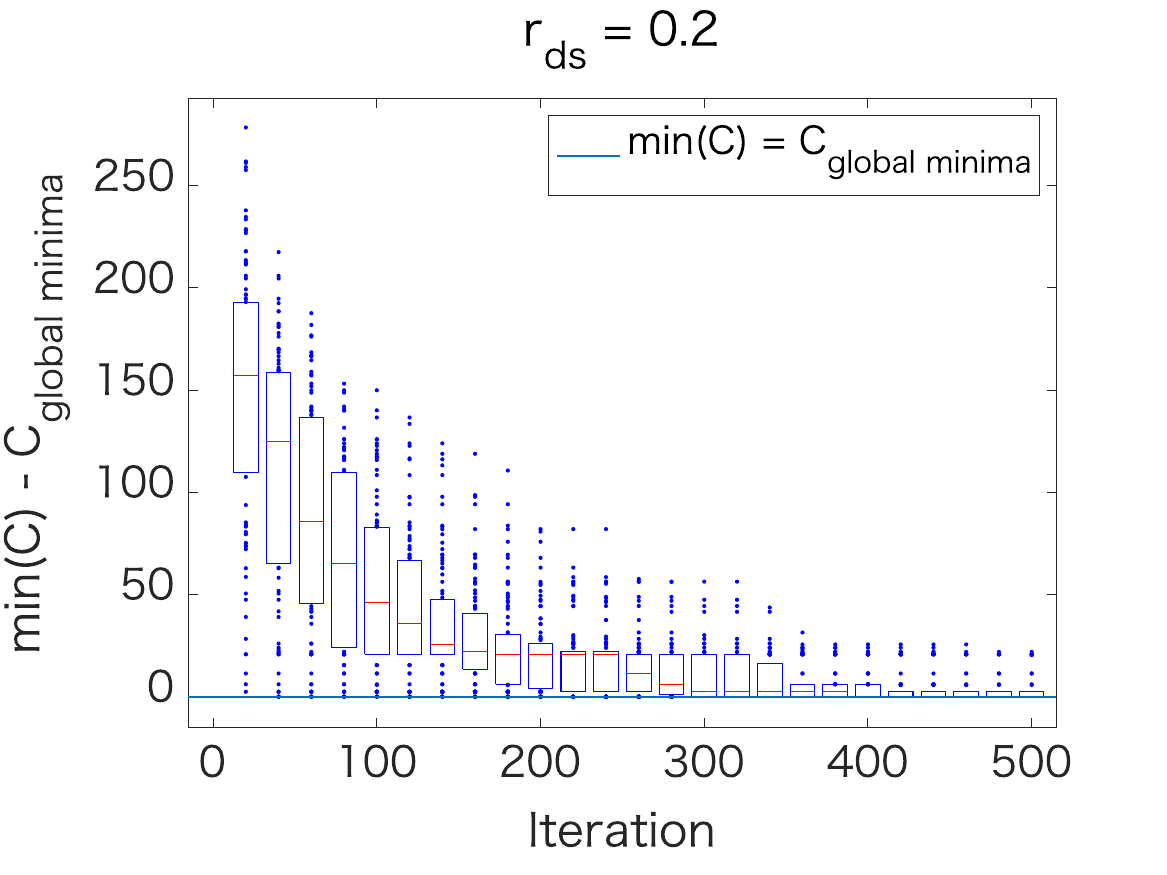}}
	\end{center}
\end{minipage}
\begin{minipage}{0.45\hsize}
	\begin{center}
		 \subfigure[]{
			 \includegraphics[width = 7cm]{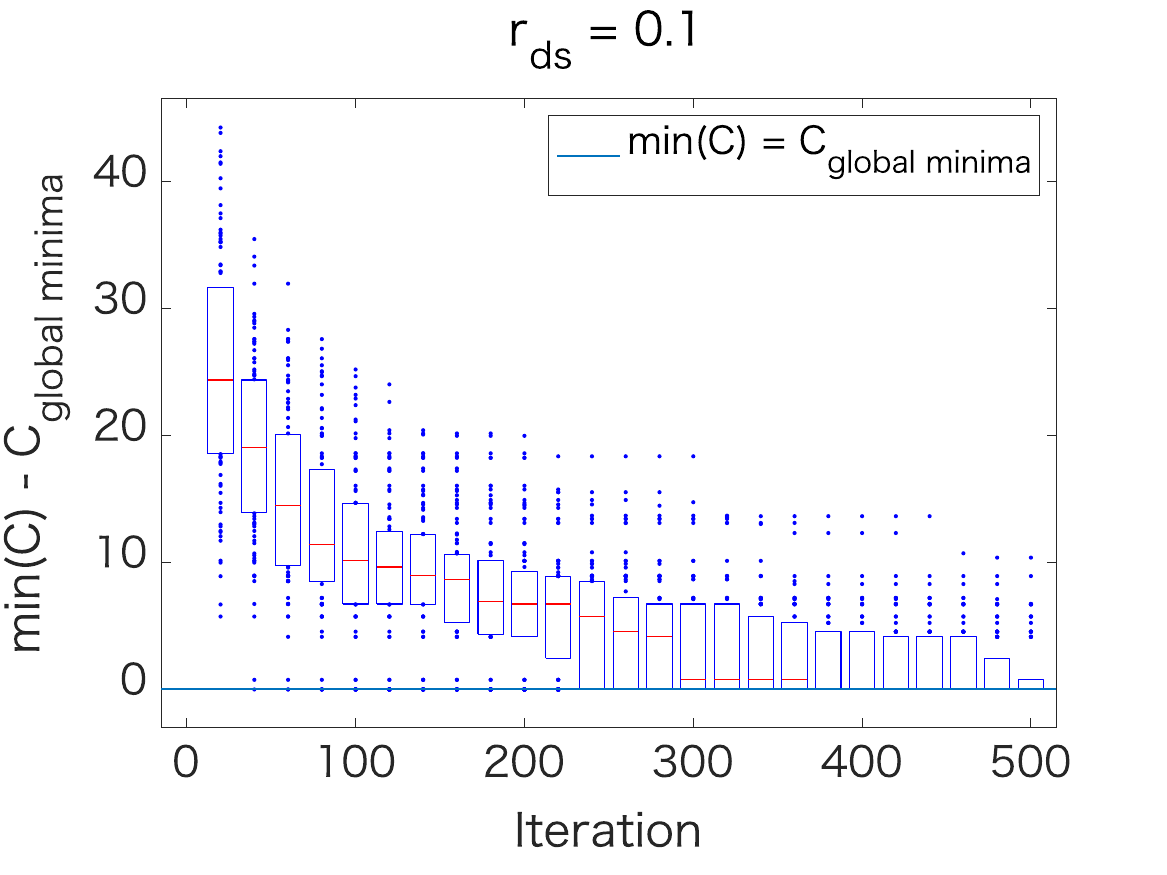}}
	\end{center}
\end{minipage}
\caption{ Performance of BO search in bin width selection for each $r_{ds}$. The predictive distribution of GP was used to compute the acquisition function. $100$ experiments were performed on the same event datasets with different initial $(\Delta_E, \Delta_{qx}, \Delta_{qy}, \Delta_{qz})$ points. The distribution of minimum values of the cost function $min(C)$ for each experiment is plotted as a function of the iteration number. Here, the baseline is set to the global minimum of the cost function $C_{\text{global minima}}$. Box plotlines show $25$th, median, and $75$th percentile from the bottom of the box to top. Whisker length is set to 0. Data points outside the boxes are plotted as dots.}
\label{fig:res_BO}
\end{figure*}

\begin{figure*}[h!]
	\begin{center}
		 \includegraphics[width = 12cm]{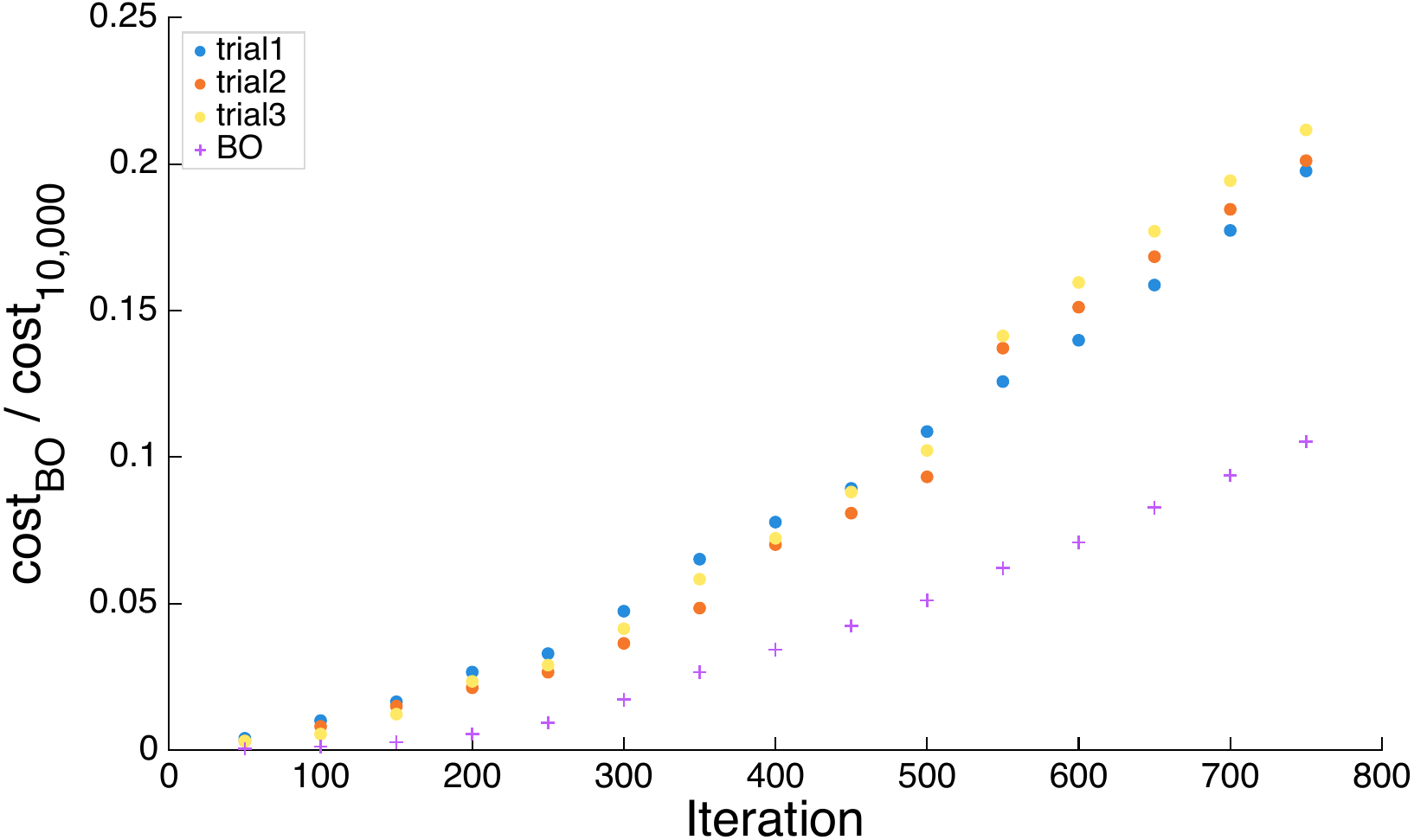}
	\end{center}
% \caption{ Comparison of computational costs of the BO search ($cost_{\rm BO}$) with that of an exhaustive search of all $10,000$ points ($cost_{\rm 10,000}$) for three sample numerical experiments by referring to the sequence of searched points. ``BO'' represents the computational cost if only BO is done. It took about 4.7 hours to conduct an exhaustive search of $10,000$ points. (CPU: 2.3 GHz Intel Xeon W: memory: 128 GB 2666 MHz DDR4). }
% [v7 Fig.6 caption] The Overleaf working copy said "3.9 hours"; restored to 3.86 h so that the caption, Results, Discussion, and Reply state the same value. Not wrapped in \rev because the reviewed manuscript (arXiv v1) already said 3.86 hours. Original working-copy line:
% \caption{ Comparison of computational costs of the BO search ($cost_{\rm BO}$) with that of an exhaustive search of all $10,000$ points ($cost_{\rm 10,000}$) for three sample numerical experiments by referring to the sequence of searched points. ``BO'' represents the computational cost of the acquisition function calculation only. trial 1-3 include both acquisition function computation time and cost function evaluation. At 500 iterations, trial 1, trial 2, and trial 3 reached 10.9\%, 9.3\%, and 10.2\% respectively. It took about 3.9 hours to conduct an exhaustive search of $10,000$ points. (CPU: Apple M1 Max; memory: 32 GB). }
\caption{ Comparison of computational costs of the BO search ($cost_{\rm BO}$) with that of an exhaustive search of all $10,000$ points ($cost_{\rm 10,000}$) for three sample numerical experiments by referring to the sequence of searched points. ``BO'' represents the computational cost of the acquisition function calculation only. trial 1-3 include both acquisition function computation time and cost function evaluation. At 500 iterations, trial 1, trial 2, and trial 3 reached 10.9\%, 9.3\%, and 10.2\% respectively. It took about 3.86 hours to conduct an exhaustive search of $10,000$ points. (CPU: Apple M1 Max; memory: 32 GB). }
\label{fig:comp_cost_rate}
\end{figure*}

\subsection{Searching for the optimal bin width by using Bayesian optimization}
We carried out numerical experiments to verify the efficiency of BO in searching for optimal bin widths in which we applied BO to the cost functions for the downsampled data. We performed $100$ experiments for each $r_{ds}$ with different initial $(\Delta_E, \Delta_{qx}, \Delta_{qy}, \Delta_{qz})$ points. We used MATLAB’s bayesopt function and set the acquisition function to EI+. The results of the numerical experiments are visualized in Fig. \ref{fig:res_BO}. Moreover, we estimated the computational complexity of the BO search and compared its computational cost with that of the $10,000$ point search.
% [v7#8 self-correction] original (the divisions-only description was inconsistent with the Fig. 6 caption):
% For estimating the computational cost, we computed the sum of the number of divisions for each ${\bf \Delta}$.
\rev{For estimating the computational cost, we combined the measured cumulative acquisition-function computation time with an estimate of the cost-function evaluation derived from the sequence of searched points.} The results are shown in Fig. \ref{fig:comp_cost_rate}.

\section{Discussion\label{sec:discussions}}
Fig. \ref{fig:cost_func} indicates that the optimal bin widths decrease as the number of data increases, which is in agreement with previous studies \cite{Shimazaki, Muto, tatsumi2022optimization}. This trend is consistent with Tatsumi et al.'s findings for Cu single crystal, confirming the general applicability of the bin-width optimization theory across different material systems. The optimal bin widths for data downsampled to $1/5$ ($r_{ds}$ = 0.2) are comparable with the resolutions determined by experimental conditions such as the sample size and the chopper timing. This implies that the measurement had some redundancy.
% [v7#9 referee(2)] original:
% Therefore, the development of a real-time termination strategy can help to prevent excessive measurement.
\rev{Therefore, the development of an online termination strategy can help to prevent excessive measurement.}

% [v7#10+#11 referee(2)] original (the opening two sentences and the real-time-performance sentence are replaced; the rest is unchanged):
% It is crucial to reduce the computational complexity of the bin-width optimization in the real-time strategy. While Tatsumi et al.\ achieved real-time optimization by performing a parallel computation on a 32-core Xeon processor \cite{tatsumi2022optimization}, our Bayesian optimization approach could search efficiently without requiring such a parallel infrastructure. [...] This significant computational cost reduction demonstrates that BO-based bin-width optimization can achieve real-time performance on standard computing environments without parallel infrastructure. [...]
\rev{It is important to reduce the computational cost of bin-width optimization for use during an experiment. Tatsumi et al.\ demonstrated rapid optimization using a parallelized Fortran implementation with MPI on a 32-core Xeon processor \cite{tatsumi2022optimization}, whereas our BO approach reduces the number of evaluated bin-width candidates without an explicit MPI-based parallel implementation.} Fig. \ref{fig:cost_func} suggests that the landscapes of the cost functions would be smooth. Since GP is effective in interpolating on smooth functions, we expect that BO will work properly. When there was a lot of data, the landscape of the cost function (Fig. \ref{fig:cost_func}) around the global minima seemed to be deep. Therefore, BO for bin-width optimization should be satisfactory in cases with a large amount of data. In fact, Fig. \ref{fig:res_BO} shows that BO is especially effective for searching for the optimal bin widths when the number of data is large. Note that we limited the bin-width search space to being a discrete one in this experiment, resulting in $min(C)$ values in the figure having a discrete structure. In our numerical experiments, it seemed reasonable to set the maximum number of iterations to $500$ (Fig. \ref{fig:res_BO}). According to Fig. \ref{fig:comp_cost_rate}, $500$ iterations of trial searches cost only about $10\%$ of the exhaustive $10,000$ points search. \rev{This reduction in evaluated search cost indicates the potential for online bin-width optimization on a workstation, although end-to-end timing remains to be measured in the intended experimental environment.} Moreover, in ``trial2'', the computational cost significantly increased for iterations between $500$ and $550$. This was due to an intensive exploration of the small bin-width area.

\rev{The dependence of the computation time on the number of collected events, $n$, differs among the processing stages. Accumulating events into the fine ``raw'' histogram requires a single pass through the event list and therefore scales as $O(n)$. After the fine histogram and summed-area table have been prepared, evaluating the cost function for a candidate bin width depends on the histogram and search-grid configuration rather than directly on $n$. We did not measure the end-to-end scaling with $n$. In an online implementation, the raw histogram could in principle be updated incrementally using only newly acquired events, although this option was not benchmarked here.}

% [v7#13 referee(2)] new paragraph: computation time vs measurement cadence (citation audit v2 sec.4)
\rev{The exhaustive search of 10,000 candidate points required approximately 3.86 h on the Apple M1 Max system used here. At 500 evaluations, the hybrid normalized computational-cost measure in Fig. \ref{fig:comp_cost_rate}, which combines measured acquisition-function time with an estimate of cost-function evaluation, was approximately 10\% of the exhaustive-search cost. This corresponds to a representative cost-equivalent estimate of approximately 23 min. The estimate was derived from the normalized computational cost rather than measured separately as an end-to-end wall-clock runtime. Combining measurements taken at multiple crystal orientations into a four-dimensional data set in momentum-transfer and energy space is an established use of time-of-flight spectroscopy \cite{Horace2016}. Reported rotation scans extend from many hours to days: a 4SEASONS study counted for 15 min at each 1$^\circ$ step, scanning from $-50^\circ$ to $180^\circ$ at 6 K and from $0^\circ$ to $105^\circ$ at 150 K \cite{MnPSe3_2024}, and a study on ARCS at the Spallation Neutron Source collected data over six days, including approximately three days over a 39$^\circ$ range at 10 K \cite{Parshall2014}. These values describe complete scans or temperature-specific data collection rather than a universal experimental time step, and an individual acquisition step can be shorter than the cost-equivalent estimate given above. A stopping check therefore need not be synchronized with every orientation: for the 15 min step cited above, evaluating the criterion at every second or third orientation gives an interval of 30--45 min, which is longer than the approximately 23 min estimate. The interval should in general be set longer than the end-to-end runtime measured on the facility computer, with an appropriate safety margin. Because the accumulation stage scales with the number of events, that runtime should be assessed at the largest event count expected in the planned measurement, or events should be accumulated incrementally so that each update processes only the newly acquired events.}

% [v7#14 referee(2)] new paragraph: reported Tatsumi et al. timings and why a direct comparison is not possible
\rev{Tatsumi et al. reported optimization times of approximately 20 s for two data sets and approximately 4 min for a third data set using their parallelized Fortran/MPI implementation on a 32-core Xeon processor \cite{tatsumi2022optimization}. A direct timing comparison is not possible because their data sets, optimization task, software implementation, hardware, and reported metric differ from those of the present study. We therefore do not claim a speed advantage over their implementation. The result demonstrated here is a reduction in evaluated search cost without an explicit MPI-based parallel implementation; a controlled same-hardware benchmark remains future work.}

% [v7#15 referee(1)] new paragraphs: initialization sensitivity, activation threshold, and initialization guidelines
\rev{Figure \ref{fig:res_BO} shows that BO convergence is sensitive to the initial points when the event count is small ($r_{ds}=0.1$). The full-data panel shows earlier concentration of the interquartile range toward the global minimum than the $r_{ds}=0.1$ panel, but the present results do not establish a universal event-count threshold above which initialization becomes irrelevant. We interpret limited data as a condition that can amplify initialization sensitivity because the cost-function minimum is less distinct at small data sizes.}

\rev{The stopping criterion is intended for use only after a sufficient amount of data has accumulated. In practice, termination decisions should therefore be disabled until a predefined minimum-data activation threshold is reached. The threshold should be selected for each experiment from pilot data and the planned acquisition schedule; a predefined fraction of the planned acquisition time is one possible operational choice, but the present study does not determine a universal fraction. At the first active evaluation, initial points should cover the discrete search region rather than be concentrated in a narrow neighborhood. In subsequent evaluations, the previous optimum provides a natural starting point. These are practical recommendations, and their performance was not separately validated in the present study.}

% [v7#16 referee(3)] new paragraph: temporal-persistence safeguard
\rev{The repeated BO searches in Fig. \ref{fig:res_BO} assess reproducibility with respect to the initial points for each fixed event data set, but they do not represent independent realizations of counting statistics or sequential stopping decisions. To reduce termination caused by transient statistical fluctuations, the stopping condition can be required to hold at $M$ consecutive evaluation times. For example, $M=3$ provides a simple temporal-persistence safeguard. The appropriate value of $M$ should be selected jointly with the evaluation interval, because increasing $M$ can suppress transient false stops but can also delay a true stop. Combined with the minimum-data activation threshold described above, this provides a conservative operational rule. Its false-stop probability and detection delay were not quantified in the present study and remain to be validated.}

\section{Conclusion\label{sec:conclusion}}
% [v7#17 referee(2)] original (full replacement; the 1/5-downsampling finding is retained in the revised text):
% To prevent redundant measurements in inelastic neutron-scattering experiments, we proposed a Bayesian-optimization-based automatic termination strategy for real-time bin-width optimization. The experiment is terminated when the optimal bin widths become smaller than the target resolutions. Through numerical experiments, we demonstrated that this Bayesian optimization can reduce the search cost to approximately 10\% of an exhaustive search. By using Bayesian optimization to efficiently search for optimal bin widths, our approach achieves real-time performance without requiring parallel computation infrastructure. Numerical experiments using Ba$_3$Fe$_2$O$_5$Cl$_2$ data confirmed that optimal bin widths decrease as data accumulates. Even the optimal bin widths for data downsampled to $1/5$ were comparable with equipment-limited resolutions, indicating measurement redundancy. This computational efficiency demonstrates that our approach can serve as a practical stopping criterion for routine inelastic neutron-scattering experiments, even on standard single-core computing environments.
\rev{To prevent redundant measurements in inelastic neutron-scattering experiments, we proposed a Bayesian-optimization-based strategy for computing a bin-width-based stopping criterion. Numerical experiments using Ba$_3$Fe$_2$O$_5$Cl$_2$ data confirmed that the optimal bin widths decrease as data accumulate, and that even the optimal bin widths for data downsampled to $1/5$ were comparable with equipment-limited resolutions, indicating measurement redundancy. Bayesian optimization reduced the evaluated search cost to approximately 10\% of that of exhaustive search. This result indicates the potential for online stopping decisions without an explicit MPI-based parallel implementation. Practical deployment requires activation only after sufficient data accumulation, an evaluation interval longer than the measured end-to-end runtime in the intended environment, and safeguards against transient stopping decisions.}

\begin{acknowledgment}
The authors thank N. Abe, T. Omi, K. Matsuura, K. Ikeuchi, R. Kajimoto, and Y. Inamura for providing us the inelastic neutron-scattering data and the experimental resolutions.
This work was partially supported by JSPS KAKENHI Grant-in-Aid for Scientific Research(A) (No. 18H04106), No. JP19H05826, and JST CREST (JPMJCR1761).
The inelastic neutron-scattering data of Ba$_3$Fe$_2$O$_5$Cl$_2$ were obtained by a chopper spectrometer 4SEASONS installed on BL01, J-PARC MLF, Japan, under the proposal No. 2016A0088.
\end{acknowledgment}

\appendix
\section{Gaussian process}
Suppose that we have a training data set $\mathcal{D}_n$ of $n$ observations, $\mathcal{D}_n = \{ ({\bf x}_i, y_i) \}_{i=1}^n$, where ${\bf x}_i$ denotes an input vector of dimension $D$, and $y_i$ denotes a scalar output. The column input vectors for all $n$ cases are aggregated in the $D \times n$ design matrix $X := ({\bf x}_1, ... , {\bf x}_n )$. We assume that each observation is the sum of the objective function $f$, and independent identically distributed Gaussian noise $\epsilon$
\begin{eqnarray}
y_i &=& f({\bf x}_i) + \epsilon, \\
\epsilon &\overset{i.i.d.}{\sim}& \mathcal{N}(0, \sigma^2).
\end{eqnarray}
We assume that the joint distribution of the observations ${\bf y}$ and a test output $f^*$ at test input ${\bf x}^*$ can be described as
\begin{eqnarray}
\left( \begin{array}{c}
{\bf y} \\
f^*
\end{array} \right) &\sim& \mathcal{N} \left(\boldsymbol{0}, \left[
\begin{array}{cc}
K(X,X) + \sigma^2I & {\bf k}(X,{\bf x}^*) \\
{\bf k}({\bf x}^*,X) & k({\bf x}^*,{\bf x}^*)
\end{array}
\label{eq:GPprior}
\right]\right).
\end{eqnarray}
Here, $k$ denotes a kernel function and $K(X,X)_{i,j} := k({\bf x}_i, {\bf x}_j)$. We can analytically calculate the distribution that $ f^*|{\bf y}$ follows.
\begin{eqnarray}
f^*|{\bf y} &\sim& \mathcal{N}(\mathrm{E}[f^*], \mathrm{V}[f^*]), \\
\mathrm{E}[f^*] &=& k(X,{\bf x}^*)^T (K(X,X) + \sigma^2I )^{-1} {\bf y}, \\
\mathrm{V}[f^*] &=& k({\bf x}^*,{\bf x}^*) - k(X,{\bf x}^*)^T (K(X,X) + \sigma^2I )^{-1} k(X,{\bf x}^*).
\end{eqnarray}

\bibliographystyle{jpsj}
\bibliography{refs}
\end{document}